\documentclass[twocolumn]{aastex631}
\usepackage{amsmath,amssymb}
\usepackage{chemformula}

\begin{document}

\title{Discovery of thionylimide, HNSO, in space: the first N-, S- and O-bearing interstellar molecule}

\author[0000-0001-9629-0257]{Miguel Sanz-Novo}
\affiliation{Centro de Astrobiolog{\'i}a (CAB), INTA-CSIC, Carretera de Ajalvir km 4, Torrej{\'o}n de Ardoz, 28850 Madrid, Spain}

\author[0000-0002-2887-5859]{V\'ictor M. Rivilla}
\affiliation{Centro de Astrobiolog{\'i}a (CAB), INTA-CSIC, Carretera de Ajalvir km 4, Torrej{\'o}n de Ardoz, 28850 Madrid, Spain}

\author[0000-0002-0183-8927]{Holger S. P. M{\"u}ller}
\affiliation{I. Physikalisches Institut, Universit{\"a}t zu K{\"o}ln, Z{\"u}lpicher Str. 77, 50937 K{\"o}ln, Germany}

\author[0000-0003-4493-8714]{Izaskun Jim\'enez-Serra}
\affiliation{Centro de Astrobiolog{\'i}a (CAB), INTA-CSIC, Carretera de Ajalvir km 4, Torrej{\'o}n de Ardoz, 28850 Madrid, Spain}

\author[0000-0003-4561-3508]{Jes\'us Mart\'in-Pintado}
\affiliation{Centro de Astrobiolog{\'i}a (CAB), INTA-CSIC, Carretera de Ajalvir km 4, Torrej{\'o}n de Ardoz, 28850 Madrid, Spain}

\author[0000-0001-8064-6394]{Laura Colzi}
\affiliation{Centro de Astrobiolog{\'i}a (CAB), INTA-CSIC, Carretera de Ajalvir km 4, Torrej{\'o}n de Ardoz, 28850 Madrid, Spain}

\author[0000-0003-3721-374X]{Shaoshan Zeng}
\affiliation{Star and Planet Formation Laboratory, Cluster for Pioneering Research, RIKEN, 2-1 Hirosawa, Wako, Saitama, 351-0198, Japan}

\author[0000-0002-6389-7172]{Andr\'es Meg\'ias}
\affiliation{Centro de Astrobiolog{\'i}a (CAB), INTA-CSIC, Carretera de Ajalvir km 4, Torrej{\'o}n de Ardoz, 28850 Madrid, Spain}

\author[0000-0001-6049-9366]{\'Alvaro L\'opez-Gallifa}
\affiliation{Centro de Astrobiolog{\'i}a (CAB), INTA-CSIC, Carretera de Ajalvir km 4, Torrej{\'o}n de Ardoz, 28850 Madrid, Spain}

\author[0000-0001-5191-2075]{Antonio Mart\'inez-Henares}
\affiliation{Centro de Astrobiolog{\'i}a (CAB), INTA-CSIC, Carretera de Ajalvir km 4, Torrej{\'o}n de Ardoz, 28850 Madrid, Spain}

\author[0000-0002-4782-5259]{Bel\'en Tercero}
\affiliation{Observatorio Astron\'omico Nacional (OAN-IGN), Calle Alfonso XII, 3, 28014 Madrid, Spain}

\author[0000-0002-5902-5005]{Pablo de Vicente}
\affiliation{Observatorio de Yebes (OY-IGN), Cerro de la Palera SN, Yebes, Guadalajara, Spain}

\author[0000-0001-7535-4397]{David San Andr\'es}
\affiliation{Centro de Astrobiolog{\'i}a (CAB), INTA-CSIC, Carretera de Ajalvir km 4, Torrej{\'o}n de Ardoz, 28850 Madrid, Spain}

\author[0000-0001-9281-2919]{Sergio Mart\'in}
\affiliation{European Southern Observatory, Alonso de C\'ordova 3107, Vitacura 763 0355, Santiago, Chile}
\affiliation{Joint ALMA Observatory, Alonso de C\'ordova 3107, Vitacura 763 0355, Santiago, Chile}

\author[0009-0009-5346-7329]{Miguel A. Requena-Torres}
\affiliation{University of Maryland, College Park, ND 20742-2421 (USA)}
\affiliation{Department of Physics, Astronomy and Geosciences, Towson University, Towson, MD 21252, USA}

\begin{abstract}

We present the first detection in space of thionylimide (HNSO) toward the Galactic Center molecular cloud G+0.693-0.027, thanks to the superb sensitivity of an ultradeep molecular line survey carried out with the Yebes 40$\,$m and IRAM 30$\,$m telescopes. This molecule is the first species detected in the interstellar medium containing, simultaneously, N, S and O. We have identified numerous $K$$_a$ = 0, 1 and 2 transitions belonging to HNSO covering from $J$$_{\rm up}$ = 2 to $J$$_{\rm up}$ = 10, including several completely unblended features. We derive a molecular column density of $N$ = (8 $\pm$ 1)$\times$10$^{13}$ cm$^{-2}$, yielding a fractional abundance relative to H$_2$ of $\sim$6$\times$10$^{-10}$, which is about $\sim$37 and $\sim$4.8 times less abundant than SO and \ch{SO2}, respectively. Although there are still many unknowns in the interstellar chemistry of NSO-bearing molecules, we propose that HNSO is likely formed through the reaction of the NSO radical and atomic H on the surface of icy grains, with alternative routes also deserving exploration. Finally, HNSO appears as a promising link between N- , S- and O- interstellar chemistry and its discovery paves the route to the detection of a new family of molecules in space.

\end{abstract}

\section{Introduction} 
\label{sec:intro}

In recent years, astrochemistry has witnessed an outburst of new interstellar detections, with more than 75 new species discovered since 2021.\footnote{https://cdms.astro.uni-koeln.de/classic/molecules} These species contain the six chemical elements essential for life: carbon (C), hydrogen (H), oxygen (O), nitrogen (N), phosphorus (P), and sulfur (S); and many of them are considered molecular precursors of prebiotic chemistry (e.g. \citealt{belloche_detection_2008,belloche2019,zeng2019,rivilla2019b,rivilla2021a,rivilla2022a,jimenez-serra2022}). However, despite the plethora of new species, NSO compounds (i.e., molecules containing simultaneously N, S and O) have attracted little attention from the astronomical community, although they appear as a promising yet unexplored link between the chemistry of N-, S- and O-bearing species in the ISM. 

On Earth, the NSO chemistry plays a key biological role in transmitting signals both within and between cells and tissues \citep{FOSTER2009,Miljkovic2013}. In particular, several [H,N,S,O] isomers (e.g. HSNO, HONS and HNSO) are thought to connect the biochemistries of two important biological messengers, nitric oxide (NO) and hydrogen sulfide (\ch{H2S}) \citep{Filipovic2012,Miljkovic2013,Ivanova2014,Nava2016,Kumar2017}. NO, as the first gasotransmitter, is involved in the regulation of vascular tone and heart function, among other physiological processes \citep{Wu2018}. Meanwhile, \ch{H2S} plays a positive role regarding antioxidative stress and inflammation regulation \citep{Zhao2024}. NSO-bearing compounds are also very abundant in oil sands \citep{Ji2021} and contain rich geological and geochemical data \citep{Quan2010,Noah2015,Ziegs2018,Chang23nso}, excelling in recording biotic and palaeoenvironmental signatures \citep{yue2023}, and thus are of interest for astrobiology.

This fact triggered us to explore the chemistry of NSO compounds in the ISM. As a proof of concept, we targeted the study of thionylimide, HNSO, 
which can be seen as an NH- analogue of sulfur dioxide, \ch{SO2}, where one O atom of O=S=O is replaced by a NH- group, yielding HN=S=O. This molecule has been suggested to be the parent species of interstellar NS \citep{barbier2006lectures}, and appears as a promising candidate since both SO and \ch{SO2} are particularly abundant in the ISM \citep{Snyder:1975cr,Cernicharo:2011gu,Vidal17}. However, despite its experimental microwave characterization being performed long ago \citep{HNSO_rot_dip_1969,HNSO_rot_1979,HNSO_rot_HFS_1993}, to our knowledge, HNSO has only been searched for toward Orion KL yielding no detection \citep{esplugues_combined_2013}. 

In this letter, we report the discovery in space of HNSO, the first interstellar N-, S- and O-bearing molecule, toward the Galactic center (GC) molecular cloud G+0.693-0.027 (hereafter G+0.693). 
We selected G+0.693 to conduct the astronomical search since it is rich in O-bearing \citep{requena-torres_organic_2006,rivilla2022a,jimenez-serra2022,SanzNovo23}, S-bearing (\citealt{rodriguez-almeida2021a,Sanz-Novo2024}), and N-bearing species \citep{zeng2018,rivilla2019b,rivilla2021b,Jimenez-Serra20,zeng2021}, of which some contain also oxygen \citep{rivilla2020b,rivilla2021a,Rivilla23,zeng2023}.

\section{Observations} 
\label{sec:obs}

We have analyzed data from an unbiased ultradeep spectral survey conducted toward the GC molecular cloud G+0.693. We covered the $Q$-band (31.075-50.424 GHz) using the Yebes 40$\,$m (Guadalajara, Spain) radiotelescope. Also, we covered three additional frequency windows with high sensitivity using the IRAM 30$\,$m (Granada, Spain) radiotelescope: 83.2$-$115.41 GHz, 132.28$-$140.39 and 142.00$-$173.81 GHz. For these observations, we used the position switching mode, centered at $\alpha$ = $\,$17$^{\rm h}$47$^{\rm m}$22$^{\rm s}$, $\delta$ = $\,-$28$^{\circ}$21$^{\prime}$27$^{\prime\prime}$, with the off position shifted by $\Delta\alpha$~=~$-885$$^{\prime\prime}$ and $\Delta\delta$~=~$290$$^{\prime\prime}$. The half power beam width (HPBW) of the Yebes 40$\,$m telescope varies between $\sim$35$^{\prime\prime}$-55$^{\prime\prime}$ (at 50 and 31 GHz, respectively; \citealt{tercero2021}) and the HPBW of the IRAM 30$\,$m radiotelescope is $\sim$14$^{\prime\prime}$$-$29$^{\prime\prime}$ across the frequency range covered. Also, we assumed that the molecular emission toward G+0.693 is extended as compared to the telescope beam \citep{Jones2012,Li2020,Zheng2024}. More details of these observations (e.g., resolution and noise levels of the spectral survey) were provided in \citet{Rivilla23} and \citet{SanzNovo23}.

\begin{center}
\begin{figure*}[ht]
     \centerline{\resizebox{1.0
     \hsize}{!}{\includegraphics[angle=0]{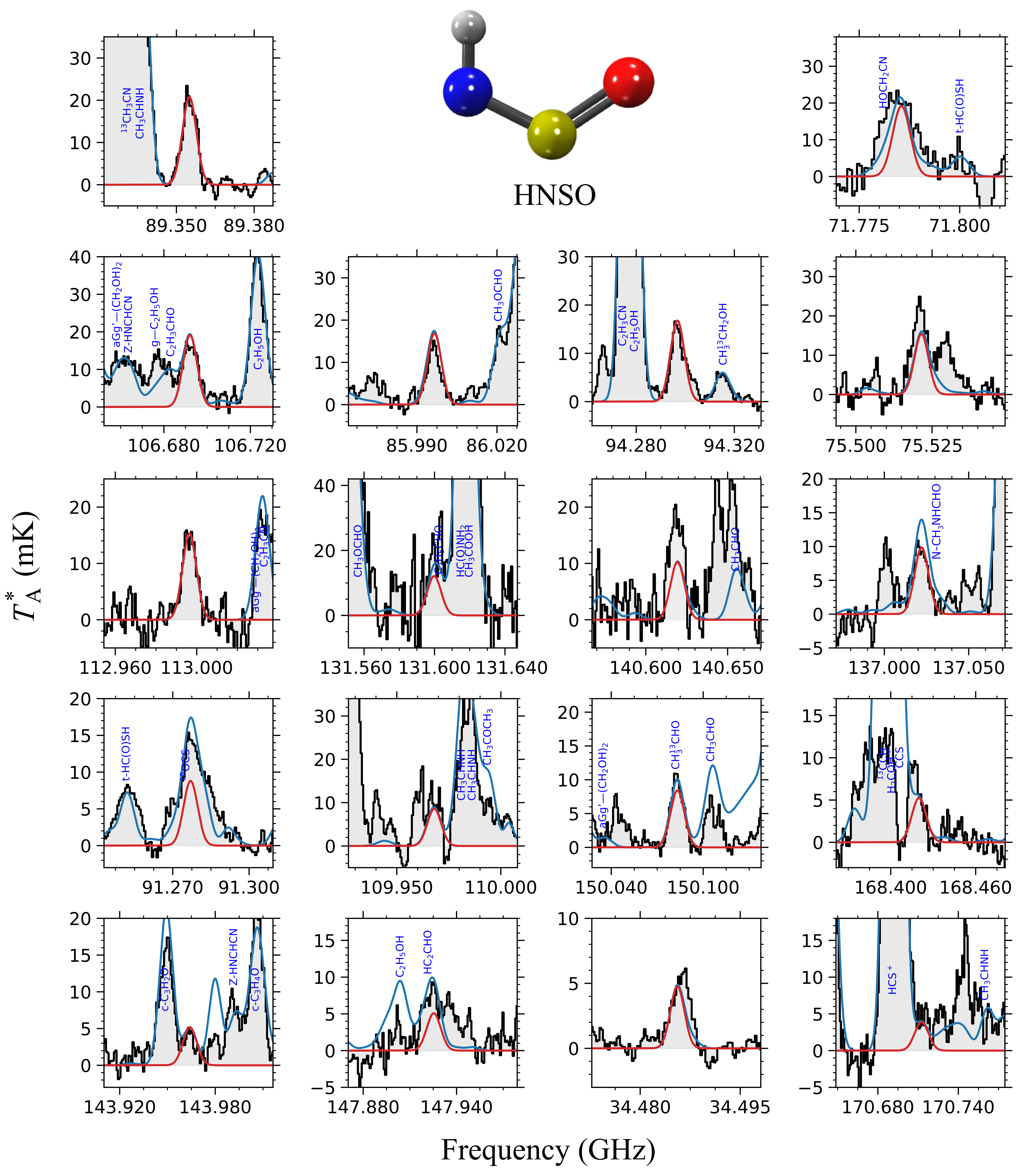}}}
     \caption{Transitions of HNSO identified toward the GC molecular cloud G+0.693–0.027 (listed in Table \ref{tab:HNSO}) sorted by decreasing peak intensity. The result of the best LTE fit of HNSO is plotted with a red line and the blue line depicts the emission from all the molecules identified to date in our survey, including HNSO (observed spectra shown as gray histograms). The structure of HNSO is also shown (nitrogen atom in  blue; oxygen atom in red, sulfur atom in yellow and hydrogen atom in white).}
\label{f:LTEspectrum}
\end{figure*}
\end{center}

\section{Detection of HNSO} 
\label{sec:detection}

HNSO may occur in the \textit{cis}- and in the \textit{trans}-configuration, but only the former (hereafter HNSO) has been detected experimentally in the laboratory. It is an asymmetric rotor close to the prolate symmetric limit with $\kappa = (2B - A - C)/(A - C) = -0.9190$, and possesses a sizable $a$-dipole moment component of 0.893~D and a much weaker $b$-component of 0.181~D. It was firstly characterized by microwave spectroscopy more than fifty years ago by \citet{HNSO_rot_dip_1969}. The astronomical line identification of HNSO has been performed using entry 63514 of the Cologne Database for Molecular Spectroscopy (CDMS\footnote{https://cdms.astro.uni-koeln.de/classic/entries/archive/HNSO/}; \citealt{Muller2005, endres2016}), explained in detail in Appendix \ref{rot_backgr}, which has been implemented into the \textsc{Madcuba} package \citep{martin2019}. We did not consider the hyperfine structure, since it is not resolved owing to the typical broad linewidths of the molecular line emission measured toward G+0.693 (FWHM $\sim$ 15$-$20 km s$^{-1}$; \citealt{requena-torres_organic_2006,requena-torres_largest_2008,zeng2018,rivilla2022c}). 

\begin{table*}
\centering
\caption{Spectroscopic information of the unblended or slightly blended transitions of HNSO detected toward G+0.693$-$0.027 (shown in Figure \ref{f:LTEspectrum}).}
\begin{tabular}{cccccccccccccc}
\hline
Frequency & Transition$^{(a)}$ & log \textit{I} (300 K) & $E$$\mathrm{_{up}}$ & rms & $\int$ $T$$\mathrm{_A^*}$d$v$ & S/N $^{(b)}$ & Blending  \\ 
(GHz) & &  (nm$^2$ MHz) & (K) &  (mK) & (mK km s$^{-1}$)  & \\
\hline
34.4854073 (9) & 2$_{1,2}$--1$_{1,1}$  & --6.1721  & 4.4 & 0.5 & 166 & 60 &  U-line \\ 
71.7851541 (19) & 4$_{0,4}$--3$_{0,3}$  & --5.1148  & 8.6 & 2.4 & 829 & 60 &  \ch{HOCH2CN}  \\
75.5211379 (19) & 4$_{1,3}$--3$_{1,2}$  & --5.1015  & 10.9 & 1.8 & 745 & 72 &  \ch{CH3CHO}  \\
85.9961329 (25) & 5$_{1,4}$--4$_{1,4}$  & --4.8860  & 14.2 & 1.1 & 342 & 54 &  Unblended* \\
89.3545678 (25) & 5$_{0,5}$--4$_{0,4}$  & --4.8337  & 12.8 & 1.0 & 477 & 83 &  Unblended* \\
91.2765634 (25) & 5$_{2,3}$--4$_{2,2}$  & --4.9013  & 20.7 & 1.1  & 547 & 87 & $^{18}$OCS \\
94.2963295 (25) & 5$_{1,4}$--4$_{1,3}$  & --4.8074  & 15.4 & 1.0 & 393 & 69 &  Unblended* \\
106.6915343 (32) & 6$_{0,6}$--5$_{0,5}$  & --4.6077  & 17.9 & 1.4 & 595 & 74 & Unblended* \\
109.9674387 (37) & 6$_{2,4}$--5$_{2,3}$  & --4.6427  & 25.9 & 1.6 & 252 & 27 &  U-line  \\
112.9958897 (33) & 6$_{1,5}$--5$_{1,4}$  & --4.5729 & 20.8 & 1.6 & 364 & 40 & Unblended*  \\
131.5991163 (43) & 7$_{1,6}$--6$_{1,5}$  & --4.3789  & 27.1 & 5.4  & 613  & 20 &  \ch{C2H5CHO} \\
137.0214017 (58) & 8$_{1,8}$--7$_{1,7}$  & --4.2898  & 31.4 & 1.1 & 278 & 44 &  N-\ch{CH3NHCHO} \\
140.6185667 (57) & 8$_{0,8}$--7$_{0,7}$  & --4.2610  & 30.5 & 1.3 & 419 & 56 &  U-line \\
143.9633403 (59) & 8$_{2,7}$--7$_{2,6}$  & --4.2782  & 38.6 & 1.3  & 180 & 24 &  Unblended* \\
147.9246044 (60) & 8$_{2,6}$--7$_{2,5}$  & --4.2550  & 39.1 & 1.0 & 313 & 55 &  \ch{HC2CHO} \\
150.0824046 (57) & 8$_{1,7}$--7$_{1,6}$  & --4.2146  & 34.2 & 1.1 & 209 & 33 & CH$_3$$^{13}$CHO \\
168.4191360 (74) & 9$_{1,8}$--8$_{1,7}$  & --4.0732  & 42.3 & 1.6 & 172 & 19 &  \ch{H2COH+}, OCS \\
170.711716 (10) & 10$_{1,10}$--9$_{1,9}$  & --4.0210  & 46.9 & 1.0 & 150 & 26 &  \ch{HCS+} \\
\hline 
\end{tabular}
\label{tab:HNSO}
\vspace*{1ex}
\tablecomments{$^{(a)}$ The rotational energy levels are labelled using the conventional notation for asymmetric tops: $J_{K_{a},K_{c}}$, where $J$ denotes the angular momentum quantum number, and the $K_{a}$ and $K_{c}$ labels are projections of $J$ along the $a$ and $c$ principal axes. $^{(b)}$ The S/N ratio is computed from the integrated signal ($\int$ $T$$\mathrm{_A^*}$d$v$) and noise level, $\sigma$ = rms $\times$ $\sqrt{\delta v \times \mathrm{FWHM}}$, where $\delta$$v$ is the velocity resolution of the spectra and the FWHM is fitted from the data. Numbers in parentheses represent the predicted uncertainty associated to the last digits.} The ``*" symbol designates those transitions used in the initial fit. We denote as U-line the line blending with a yet unidentified feature.
\end{table*}

We used the Spectral Line Identification and Modeling (SLIM) tool (version from 2023, November 15) within \textsc{Madcuba}, which, assuming a Local Thermodynamic Equilibrium (LTE) excitation condition, enables the creation of LTE synthetic spectra to be subsequently compared with the observed astronomical data. After evaluating the emission of more than 130 molecules previously identified toward G+0.693, we managed to detect numerous $K$$_a$ = 0, 1 and 2 transitions spanning from $J$$_{\rm up}$  = 2 to $J$$_{\rm up}$  = 10. The most intense unblended or slightly blended features are shown in Figure \ref{f:LTEspectrum}, and their spectroscopic information is listed in Table \ref{tab:HNSO}. Among them, we found six completely unblended transitions. Other transitions reproduce well the observations, once the blending with emission from other molecules previously detected toward G+0.693 is considered (see Figure \ref{f:LTEspectrum} and Table \ref{tab:HNSO}). The remaining lines, blended with more prominent transitions from other species, are also in agreement with the observed spectra.

The best LTE modelling for HNSO was achieved using a two-step approach as described in San Andr\'es et al. (submitted, 2024). The line width (FWHM) was constrained first using exclusively the aforementioned unblended transitions (marked with a ``*" symbol in Table \ref{tab:HNSO}), obtaining a value of FWHM = 21.5 $\pm$ 0.6 km s$^{-1}$ in an initial nonlinear least-squares LTE fit using the \textsc{Autofit} tool within SLIM \citep{martin2019}. Afterward, we performed a second fit that includes all the transitions shown in Figure \ref{f:LTEspectrum} and Table \ref{tab:HNSO}, with the exception of the 8$_{0,8}$--7$_{0,7}$ (slightly contaminated with an unidentified line), fixing the line width and leaving as free parameters the excitation temperature, $T_{\rm ex}$, radial velocity, $v$$_{\rm LSR}$, and column density, $N$. We derived a molecular column density of $N$(HNSO) = (8 $\pm$ 1) $\times$10$^{13}$ cm$^{-2}$, which yields a fractional abundance with respect to molecular hydrogen of (6 $\pm$ 1) $\times$ 10$^{-10}$, using $N$(H$_{2}$) = 1.35$\times$10$^{23}$ cm$^{-2}$ as derived by \citet{martin_tracing_2008} employing C$^{18}$O as a total H$_2$ column density tracer and assuming a C$^{18}$O/H$_2$ abundance ratio of 1.7$\times$10$^{7}$, \citet{frerking_relationship_1982}). Additionally, we obtained a $T_{\rm ex}$ = 11 $\pm$ 2 K and a $v$$_{\rm LSR}$ = 68 $\pm$ 2 km s$^{-1}$, which are consistent with those found for other molecular species toward the same source (see e.g. \citealt{requena-torres_organic_2006,zeng2018}). As shown in Table \ref{tab:HNSO}, the detected transitions cover a wide range of $E_{\rm up}$, which allowed us to derive an accurate estimate on the $T_{\rm ex}$. The fitted line profiles of HNSO are shown using a red line in Figure \ref{f:LTEspectrum} overlaid with the observed spectra (in gray). In blue, we report the predicted spectrum considering all molecular species identified and analyzed toward G+0.693. 

\begin{figure}
\centerline{\resizebox{0.95\hsize}{!}{\includegraphics[angle=0]{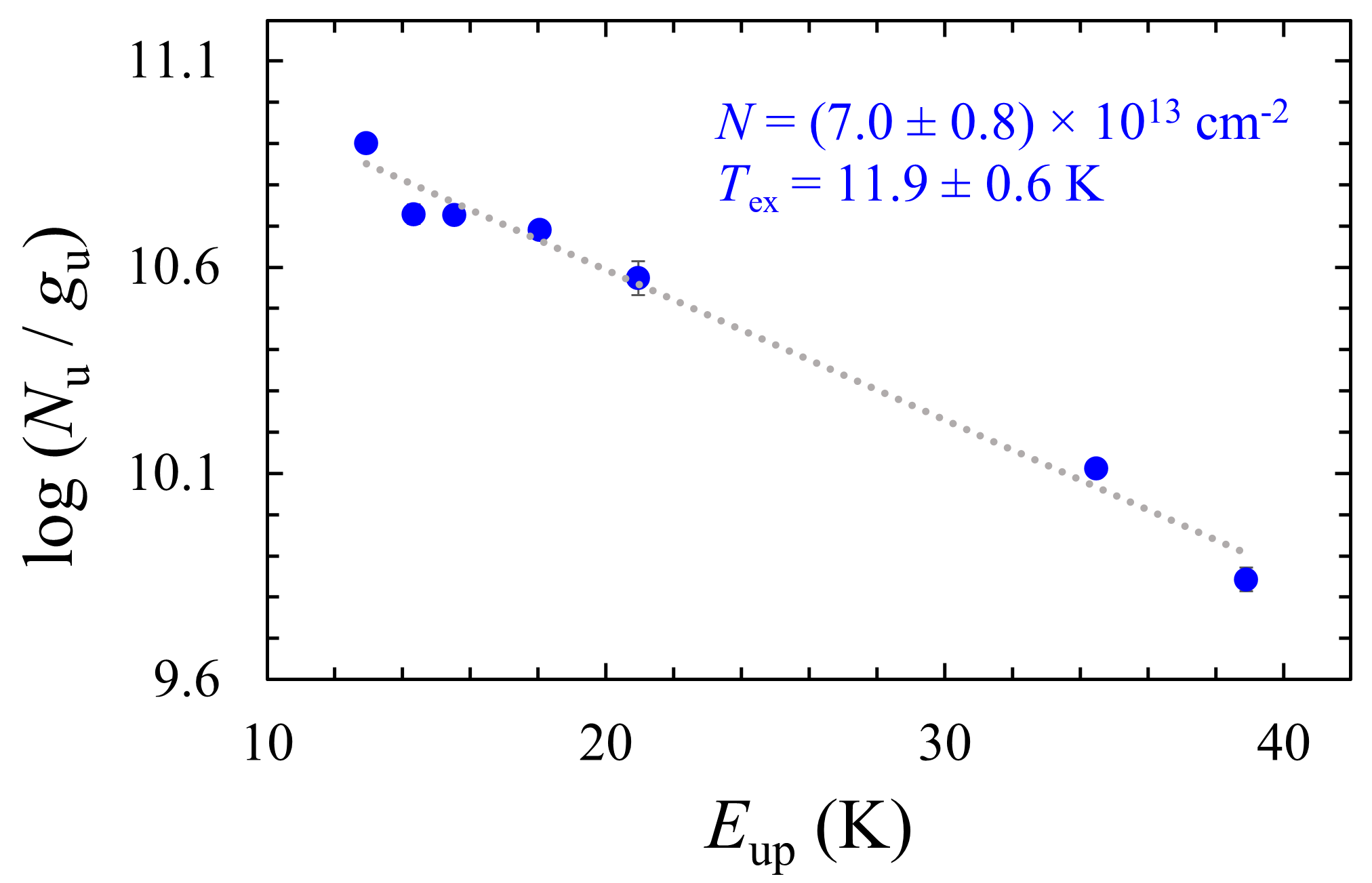}}}
\caption{Rotational diagram of HNSO toward G+0.693 (blue dots, including 1$\sigma$ errors). The best linear fit to the data points is depicted using a gray dotted line. The derived values for the molecular column density, $N$, and the excitation temperature, $T_{\rm ex}$, are shown in blue.} 
\label{f:rotdiagram}
\end{figure}

We have also performed a complementary rotational diagram analysis \citep{goldsmith1999}, as implemented in \textsc{Madcuba}, using transitions that are unblended with the emission from other molecules, together with the 8$_{1,7}$--7$_{1,6}$ transition (at 150.0824046 GHz) which is negligibly contaminated by the emission of CH$_3$$^{13}$CHO, and considering the velocity-integrated intensity over the line width \citep{rivilla2021a}. We obtained the following physical parameters for HNSO: $N$ = (7.0 $\pm$ 0.8) $\times$10$^{13}$ cm$^{-2}$ and $T_{\rm ex}$ = 11.9 $\pm$ 0.6 K, which agree well with those derived from the \textsc{AUTOFIT}. The results of the analysis are depicted in Figure \ref{f:rotdiagram}. 

\begin{table*}
\centering
\caption{Derived physical parameters for relevant N-, S- and/or O-bearing molecules related to HNSO detected toward G+0.693-0.027.}
\begin{tabular}{ c c c c c c c  c}
\hline
\hline
 Molecule & Formula & $N$   &  $T_{\rm ex}$ & $v$$_{\rm LSR}$ & FWHM  & Abundance$^a$ & Ref.$^b$   \\
 & & ($\times$10$^{14}$ cm$^{-2}$) & (K) & (km s$^{-1}$) & (km s$^{-1}$) & ($\times$10$^{-10}$)    \\
\hline
Thionylimide & HNSO & 0.8 $\pm$ 0.1 & 11 $\pm$ 2  &  68 $\pm$ 2  & 21.5$^c$  & 6 $\pm$ 1 &  (1) \\
Sulfur monoxide ($^{18}$O isotopologue) & S$^{18}$O & 0.120 $\pm$ 0.003 & 6.9$^c$  & 67.9 $\pm$ 0.3 & 24.8 $\pm$ 0.7 & 0.089 $\pm$ 0.008 & (2) \\
Sulfur monoxide & SO & 30.06 $\pm$ 0.07$^d$ & - & - & - & 224 $\pm$ 20 & (2) \\
Sulfur dioxide & SO$_2$ & 3.84 $\pm$ 0.05 & 18.9 $\pm$ 0.2 & 68.9 $\pm$ 0.1 & 21.1 $\pm$ 0.3 & 28 $\pm$ 3 & (1) \\
Nitrogen sulfide & NS & 2.8 $\pm$ 0.3 & 7.5 $\pm$ 0.5 & 70.6 $\pm$ 0.6 & 19 $\pm$ 1 & 21 $\pm$ 2 & (1) \\
Isothiocyanic acid & HNCS & 0.62 $\pm$ 0.01 & 20.4 $\pm$ 0.5 & 66.7 $\pm$ 0.3 & 21.0$^c$ & 4.6 $\pm$ 0.4 & (3) \\
Isocyanic acid ($K$$_a$ = 0)& HNCO & 32 $\pm$ 1 & 17 $\pm$ 1 & 67 $\pm$ 1 & 23 $\pm$ 1 & 239 $\pm$ 21 &  (4) \\
Cyanic acid & HOCN & 0.213 $\pm$ 0.004 & 7.4 $\pm$ 0.2 & 68 $\pm$ 0.2 & 19.2 $\pm$ 0.3 & 0.16 $\pm$ 0.01 &  (5) \\
\hline
\end{tabular}
\label{tab:comparison}
\vspace{0mm}
\vspace*{1ex}
\tablecomments{$^a$ We adopted $N_{\rm H_2}$ = 1.35$\times$10$^{23}$ cm$^{-2}$, from \citet{martin_tracing_2008}, assuming an uncertainty of 15\% of its value. $^b$ References: (1) This work; (2) \citet{rivilla2022b}; (3) \citet{Sanz-Novo2024}; (4) \citet{zeng2018}; (5) \citet{rivilla2022c}. $^c$ Value fixed in the fit. $^d$ Computed using the most optically thin isotopologue S$^{18}$O and assuming a $^{16}$O/$^{18}$O = 250 in the CMZ \citep{wilson_abundances_1994}.}
\label{tab:g0693}
\end{table*}

\section{Discussion} 
\label{sec:disc}

To understand how HNSO fits in a broader astrochemical context, we can compare its abundance with that reported for plausible precursors, such as SO \citep{rivilla2022b} and NS (the analysis is detailed in Appendix \ref{AnalysisNS}), as well as other structurally related molecules also detected toward G+0.693 (see Table \ref{tab:comparison}) highlightling \ch{SO2} (analyzed in Appendix \ref{AnalysisSO2}), HNCS \citep{Sanz-Novo2024}, HNCO \citep{zeng2018} and HOCN \citep{rivilla2022c}. We found a $N$(SO)/$N$(HNSO) ratio of 37 $\pm$ 4 toward G+0.693, which is significantly lower than the lower limit ratio derived toward Orion KL ($N$(SO)/$N$(HNSO) $\geq$ 2600; \citealt{esplugues_combined_2013}). Moreover, we obtained a $N$(\ch{SO2})/$N$(HNSO) ratio of 4.8 $\pm$ 0.6, a $N$(\ch{HNCS})/$N$(HNSO) ratio of 0.8 $\pm$ 0.1, a $N$(HNCO)/$N$(HNSO) ratio of 40 $\pm$ 6 and a $N$(HOCN)/$N$(HNSO) ratio of 0.27 $\pm$ 0.4. Consequently, HNSO appears to be a rather abundant molecule within interstellar S- chemistry, despite having gone unnoticed until now. Our observations are in line with the results presented in \citet{Sanz-Novo2024} for a variety of well-known interstellar S-bearing molecules, which suggest that S is not significantly depleted toward G+0.693 compared to other astronomical sources \citep{martin-domenech16,Vidal17,Marcelino23,Fuente23}.

\begin{figure}
\centerline{\resizebox{1\hsize}{!}{\includegraphics[angle=0]{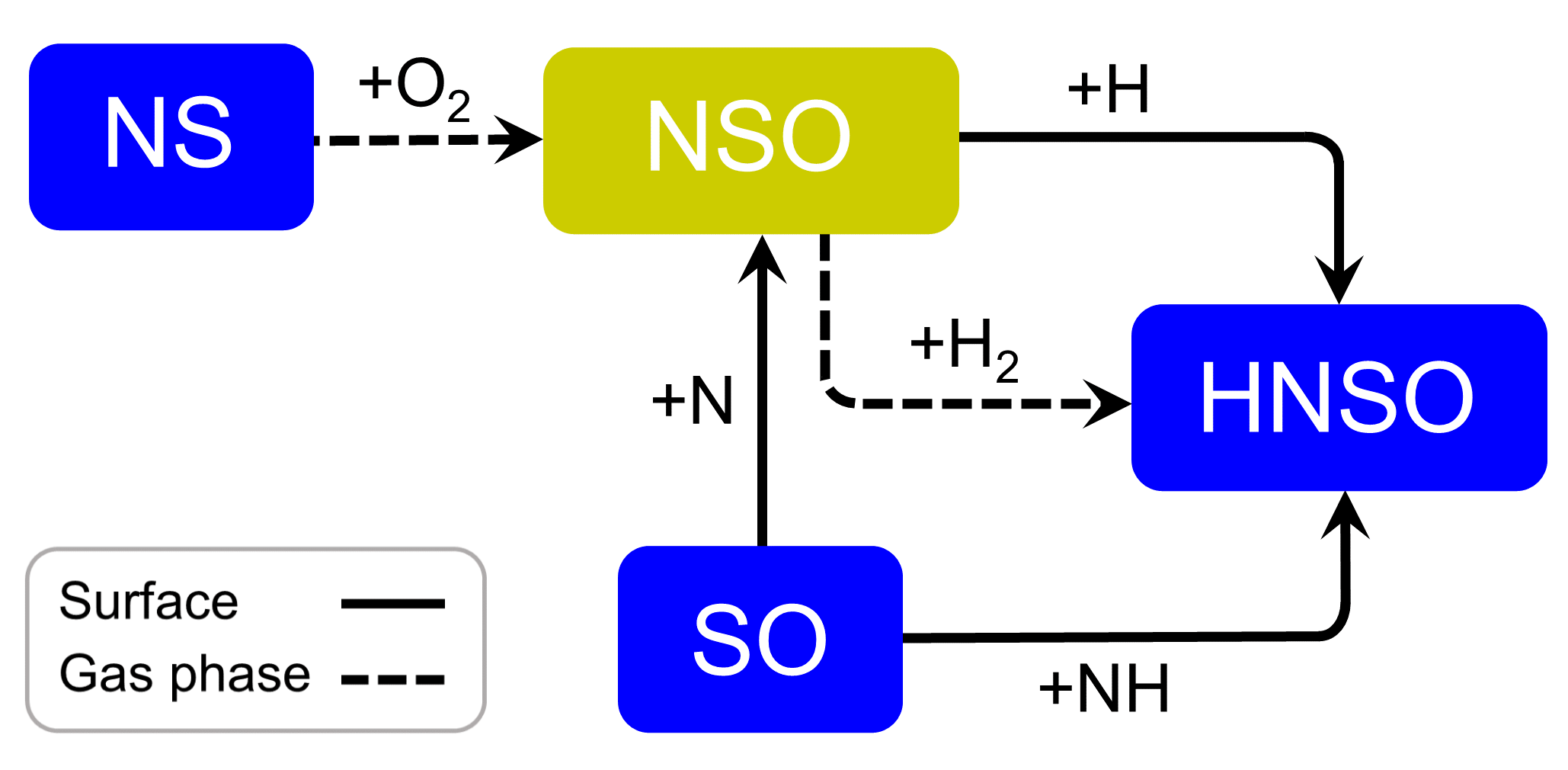}}}
\caption{Suggested chemical routes for the formation of \ch{HNSO} in the ISM. We show in blue molecules that have been identified toward G+0.693 and in yellow molecules that have not been searched for toward G+0.693 because spectroscopy is not available (NSO). Surface reactions are shown in black solid lines while gas-phase reactions are depicted with dashed arrows.}
\label{f:model}
\end{figure}

Although HNSO could be a link between N-, S- and O- interstellar chemistry, this species is missing in commonly-used chemical networks, such as the Kinetic Database for Astrochemistry (KIDA; \url{http://kida.astrophy.u-bordeaux.fr/}) or in the UMIST Database for Astrochemistry (UDfA; \citealt{Millar2024}). Therefore, further effort is required to account for the formation of NSO-compounds, which will contribute to achieve a better agreement between observations and chemical models of S-, N- and O-bearing species.

Concerning its production in the ISM, HNSO is most likely formed through the grain-surface reaction between the NSO radical --whose rotational signatures remain unknown and, therefore, it has not been searched for in the ISM so far-- and atomic hydrogen. This hydrogenation process is feasible at the dust temperatures measured in the Galactic Center ($\sim$20 K; \citealt{rodriguez-fernandez2000,Etxaluze2013}). Once HNSO is formed on the grains, it can be subsequently transferred to the gas phase thanks to shock-induced desorption. Alternatively, if we assume a similar grain-surface formation pathway as that explored for HNCO by \citet{Quenard2018}, i.e., NH + CO  $\rightarrow$ HNCO \citep{fedoseev2016}, we can propose the following alternative radical-radical diffusion reaction: NH + SO $\rightarrow$ HNSO. Note, however, that the diffusion of radicals on grain surfaces is not expected to occur efficiently at temperatures below 30 K \citep{Garrod:2008tk,garrod2013ApJ}, and thus this mechanism is not expected to be dominant at the low dust temperatures measured in the Galactic Center. Nonetheless, as recently shown in the models of \citet{Garrod2022}, non-diffusive chemistry could alternatively yield significant amounts of HNCO even at low dust temperatures via the reaction NH + CO $\rightarrow$ HNCO, which suggests that the same could apply to the reaction NH + SO $\rightarrow$ HNSO. 

In the gas phase, the hydrogenation process has been explored theoretically using high-level coupled cluster methodologies for all possible NSO radical isomers \citep{Kumar2017}, suggesting that the formation of their corresponding [H,N,S,O] hydrides (i.e., HNSO, HSNO and SNOH) is exothermic and barrierless, being thus feasible under interstellar conditions. Nevertheless, these authors showed that, once formed, these hydrides will likely decompose yielding SH and NO, SN and OH or SO and NH radicals. Another possible gas-phase route is the reaction between NSO and H$_2$, yielding HNSO + H, which, to the best of our knowledge, remains unexplored.

On the other hand, the formation of the NSO radical, a mixed oxide of nitrogen and sulfur, and also the most stable member of the [N,S,O] isomeric family \citep{Kumar2017}, remains uncharted as well. On grains, a possible diffusion mechanism of atomic N on the dust-grain surface cannot be a priori discarded, which, if efficient enough at the typical low temperatures of the icy grain mantles, could react with SO yielding NSO. In this context, it has recently been shown that the diffusion of atomic C is feasible at temperatures above 22 K on icy grain surfaces \citep{Tsuge2023}, which is on the order of the dust temperature of the Galactic Center. Consequently, it would be interesting to extend this investigation on plausible N-insertion reactions driven by the diffusion of N atoms on interstellar ices. Conversely, non-diffussive grain-surface reactions between N and SO may also play a role if they are formed \textit{in situ} \citep{Garrod2022} (e.g., through early photochemistry within the ices), subsequently yielding NSO. On a similar note, the reaction N(gas) + SO (grain) forming NSO on the grain surface could also be studied, since C-atom addition reactions between C coming from gas-phase and grain-surface species are proved to be efficient \citep{Fedoseev2022}.

In addition, we propose the following gas-phase reaction: NS + \ch{O2} $\rightarrow$ NSO + O, similarly to the NCO radical, which is produced via CN + \ch{O2}  $\rightarrow$ NCO + O  (reaction rate of 2.4 $\times$ 10$^{-11}$ cm$^{-3}$ s$^{-1}$; \citealt{Glarborg1998,Marcelino:2018cp}), given that NS is also detected toward G+0.693 (see Appendix \ref{AnalysisNS}). We find a molecular column density of $N$(NS) = (2.8 $\pm$ 0.3) $\times$ 10$^{14}$ cm$^{-2}$, which indicates that it is $\sim$3.5 times more abundant than HNSO, suggesting that the route starting with NS as plausible precursor may be relevant only if most of the reservoir of NS was locked up in the formation of HNSO. Nonetheless, in this case the NSO radical would need to be subsequently depleted in order to react with atomic H on the ice. All in all, NSO appears as a promising molecule to be studied by the laboratory spectroscopic community.

The feasibility of all the aforementioned routes also needs to be further explored, both theoretically and in the laboratory, to provide conclusive clues on the formation pathways of HNSO. Particularly, the spectroscopic study of other [H,N,S,O] isomers, highlighting the \textit{trans}- form of HNSO, certainly merits attention, given that several steroisomers\footnote{Isomers that possess identical constitution, but differ in the three-dimensional orientations of their atoms.} have been detected toward G+0.693 (e.g., \textit{cis}- and \textit{trans}-formic acid, HCOOH, and the high-energy \textit{cis}-conformer of carbonic acid, HOCOOH; \citealt{SanzNovo23}). 
 
The results presented in this letter confirm that G+0.693 is an astrochemical niche for the detection of new S-bearing species, even combined with O and/or N. This fact opens the door for the investigation of a new family of interstellar molecules. Moreover, the discovery of the first interstellar NSO-bearing species shall help in the identification of hitherto unidentified organo-sulfur molecules in the gas phase, pushing also the frontiers of known chemical complexity in the ISM and its possible contribution to prebiotic chemistry.

\software{1) Madrid Data Cube Analysis (\textsc{Madcuba}) on ImageJ is a software developed at the Center of Astrobiology (CAB) in Madrid; \url{https://cab.inta-csic.es/madcuba/}; \citet{martin2019}; version from 2023 November 15.}

\begin{acknowledgments}
 
We are grateful to the IRAM 30$\,$m and Yebes 40$\,$m telescopes staff for their help during the different observing runs, highlighting project 21A014 (PI: Rivilla), project 018-19 (PI: Rivilla) and project 123-22 (PI: Jim\'enez-Serra). The 40$\,$m radio telescope at Yebes Observatory is operated by the Spanish Geographic Institute (IGN, Ministerio de Transportes, Movilidad y Agenda Urbana). IRAM is supported by INSU/CNRS (France), MPG (Germany) and IGN (Spain). M. S. N. acknowledges a Juan de la Cierva Postdoctoral Fellow proyect JDC2022-048934-I, funded by the Spanish Ministry of Science, Innovation and Universities/State Agency of Research MICIU/AEI/10.13039/501100011033 and by the European Union “NextGenerationEU”/PRTR”. V. M. R.  acknowledges support from project number RYC2020-029387-I funded by MCIN/AEI/10.13039/501100011033 and by "ESF, Investing in your future", and from the Consejo Superior de Investigaciones Cient{\'i}ficas (CSIC) and the Centro de Astrobiolog{\'i}a (CAB) through the project 20225AT015 (Proyectos intramurales especiales del CSIC).. I. J.-S., J. M. -P., L. C, A. M., and A. M. H. acknowledge funding from grants No. PID2019-105552RB-C41 and PID2022-136814NB-I00 from MICIU/AEI/10.13039/501100011033 and by “ERDF A way of making Europe”. A. M. has received support from grant PRE2019-091471, funded by MICIU/AEI/10.13039/501100011033 and by 'ESF, Investing in your future'. A. M. H. acknowledges funds from Grant MDM-2017-0737 Unidad de Excelencia “Mar{\'i}a de Maeztu" Centro de Astrobiolog{\'i}a (CAB, INTA-CSIC). DSA also extends his gratitude for the financial support provided by the Comunidad de Madrid through the Grant PIPF-2022/TEC-25475. P. dV. and B. T. thank the support from MICIU through project PID2019-107115GB-C21. B. T. also thanks the Spanish MICIU for funding support from grant PID2022-137980NB-I00. H. S. P. M. thanks Evan Robertson and Don McNaughton for providing the infrared GSCDs. He also acknowledges support by the Deutsche Forschungsgemeinschaft (DFG) through the collaborative research grant SFB~1601 (project ID 500700252), subprojects Inf and A4. S. Z. acknowledge the support by RIKEN Special Postdoctoral Researchers Program.

\end{acknowledgments}

\bibliography{rivilla,bibliography}{}
\bibliographystyle{aasjournal}

\newpage
\appendix
\twocolumngrid

\restartappendixnumbering
 
\section{Spectroscopic properties and parameters of thionylimide}
\label{rot_backgr}

The rotational spectrum of thionylimide, HNSO, was investigated by conventional microwave spectroscopy \citep{HNSO_rot_dip_1969,HNSO_rot_1979} with accuracies of 100 and 50~kHz, respectively. A Fourier transform microwave (FTMW) spectroscopy was capable of resolving not only the hyperfine structure (HFS) splitting of $^{14}$N, but also that of $^1$H \citep{HNSO_rot_HFS_1993}. The ground state rotational parameters were improved later through ground state combination differences (GSCDs) from a high-resolution infrared study \citep{HNSO_IR-2_gs-par_1996}.

We have fitted the available rotational data by applying the rotational and centrifugal distortion parameters from \citet{HNSO_IR-2_gs-par_1996} and the HFS parameters from \citet{HNSO_rot_HFS_1993}. Moreover, the rotational spectrum calculated from the resulting spectroscopic parameters displayed fairly large uncertainties already for transitions with low quantum numbers. The underlying GSCDs from \citet{HNSO_IR-2_gs-par_1996} were unfortunately not available anymore from the corresponding author of that study. We received GSCDs from the authors of a different high-resolution infrared study \citep{HNSO_IR_2006}. Inclusion of essentially all of these GSCDs improved the uncertainties of the rotational and centrifugal distortion parameters greatly, permitted the sextic distortion parameters determined in the earlier infrared study and even the octic $L_K$ to be fit as well. Only $L_{KKJ}$ was very uncertain and was kept fixed to the value from the earlier study \citep{HNSO_IR-2_gs-par_1996}. Since the molecule is rather close to the prolate symmetric limit, it is advisable to employ Watson's S reduction of the rotational Hamiltonian instead of the A reduction as used earlier. The change in reduction had the advantage that the parameter $h_3$, which is the equivalent to $\phi _K$ in the A reduction, could be omitted in the present fit without significant deterioration of the quality of the fit.

The $^1$H HFS splitting of 3.0~kHz observed in the $F_1 = 2 - 1$ component of the $J = 1 - 0$ $a$-type transition \citep{HNSO_rot_HFS_1993} could not be explained by $^1$H nuclear spin-rotation coupling, but required the $^1$H-$^{14}$N nuclear spin-nuclear spin coupling to be considered as well. This coupling is described by two terms of which the direct term can be calculated well from the geometry for light nuclei. Deviations are small even for moderately heavy nuclei, such as Cl, as are the contributions from the indirect term, but matter both for heavier nuclei, as shown for example in a study on BrF and IF \citep{BrF_IF_rot_1995}. The HN bond is almost perfectly aligned with the $b$-axis, and a value of $S_{bb} = -8.02$~kHz was derived from the structure. The $^1$H and $^{14}$N nuclear spin-rotation parameter $C_{ii}$(H) and $C_{ii}$(N) were necessary in the fit, but were not determined very accurately; the uncertainties were in part as large as the values, and in the most favorable cases smaller by only by a factor of a few. Therefore, we evaluated these parameters through quantum-chemical calculations employing the commercially available program Gaussian~16 \citep{Gaussian16C}. We performed B3LYP hybrid density functional \citep{Becke_1993,LYP_1988} calculations using the aug-cc-pwCVTZ basis set (i.e., a valence basis set of triple zeta quality and augmented with diffuse basis functions \citealt{cc-pVXZ_1989} and with core-correlating basis functions \citealt{core-corr_2002}). Trial fits with these parameters kept fixed or with one to all released gave fits of essentially the same quality. Also, none of the parameters released in the fit changed its value significantly outside the respective parameters. Therefore, we kept the nuclear spin rotation parameters fixed at the quantum-chemically calculated values. The resulting spectroscopic parameters are given in Table~\ref{HNSO-spec-parameters} together with previous values \citep{HNSO_IR-2_gs-par_1996,HNSO_rot_HFS_1993}. The rotational spectrum calculated from these parameters is available in the catalog section of the CDMS (entry 63514). Its accuracy is sufficient for radio astronomical observations of cold objects, where molecules show low $T_{\rm ex}$ (i.e., 10-20 K), including prestellar cores and also the G+0.693 molecular cloud, and probably also for a warm source with $T_{\rm ex}$ $\lesssim$ 80$~$K. However, this catalogue may not be accurate enough for, e.g., hot cores or hot corinos with $T_{\rm ex}$ $\gtrsim$ 100$~$K. The underlying line, parameter, and fit files along with other auxiliary files are available in the catalog archive of the CDMS \citep{Muller2005, endres2016}. 


\begin{table*}
  \caption{Present and previous spectroscopic parameters$^a$ (MHz) of thionylimide.}
  \label{HNSO-spec-parameters}
  \begin{tabular}{lr@{}lr@{}ll}
  \hline
   Parameter                & \multicolumn{2}{c}{Present} & \multicolumn{2}{c}{Previous} & Parameter                  \\
  \hline
$A$                         &  49315&.8592~(84)           &  49315&.8695~(36)           & $A$                         \\
$B$                         &   9869&.75936~(23)          &   9869&.80628~(24)          & $B$                         \\
$C$                         &   8205&.14749~(23)          &   8205&.10205~(24)          & $C$                         \\
$D_K$                       &      1&.46953~(29)          &      1&.47025~(3)           & $\Delta_K$                  \\
$D_{JK} \times 10^3$        &  $-$89&.408~(41)            &  $-$90&.746~(8)             & $\Delta_{JK} \times 10^3$   \\
$D_J \times 10^3$           &      6&.6754~(21)           &      6&.9107~(6)            & $\Delta_J \times 10^3$      \\
$d_1 \times 10^3$           &   $-$1&.9211~(10)           &      1&.9227~(2)            & $\delta_J \times 10^3$      \\
$d_2 \times 10^3$           &   $-$0&.11925~(42)          &     23&.122~(16)            & $\delta_K \times 10^3$      \\
$H_K \times 10^6$           &    143&.2~(11)              &    143&.5~(1)               & $\Phi_K \times 10^6$        \\
$H_{KJ} \times 10^6$        &   $-$9&.20~(11)             &  $-$10&.42~(7)              & $\Phi_{KJ} \times 10^6$     \\
$H_{JK} \times 10^6$        &   $-$0&.317~(18)            &      0&.055~(20)            & $\Phi_{JK} \times 10^6$     \\
$H_J \times 10^9$           &     11&.92~(67)             &     12&.49~(13)             & $\Phi_J \times 10^9$        \\
$h_1 \times 10^9$           &      5&.86~(38)             &      6&.72~(8)              & $\phi_J \times 10^9$        \\
$h_3 \times 10^9$           &       &$-$                  &     12&.1~(6)               & $\phi_K \times 10^6$        \\
$L_{K}   \times 10^9$       &  $-$16&.0~(11)              &  $-$15&.90~(7)              & $L_{K}   \times 10^9$       \\
$L_{KKJ} \times 10^9$       &      1&.1                   &      1&.10~(4)              & $L_{KKJ} \times 10^9$       \\
$\chi _{aa}$                &   $-$1&.5757~(16)           &   $-$1&.5756~(29)           & $\chi _{aa}$                \\
$\chi _{bb}$                &   $-$0&.0250~(14)           &   $-$0&.0255~(24)$^b$       & $\chi _{bb}$                \\
$\chi _{cc}$                &      1&.6007~(14)$^b$       &      1&.6011~(24)$^b$       & $\chi _{cc}$                \\
$C_{aa}(\rm N) \times 10^3$ &      9&.38                  &       &$-$                  & $C_{aa}(\rm N) \times 10^3$ \\
$C_{bb}(\rm N) \times 10^3$ &      2&.60                  &      2&.3~(15)$^b$          & $C_{bb}(\rm N) \times 10^3$ \\
$C_{cc}(\rm N) \times 10^3$ &      0&.65                  &      0&.8~(15)$^b$          & $C_{cc}(\rm N) \times 10^3$ \\
$C_{aa}(\rm H) \times 10^3$ &   $-$4&.44                  &       &$-$                  & $C_{aa}(\rm H) \times 10^3$ \\
$C_{bb}(\rm H) \times 10^3$ &      1&.62                  &      1&.9~(35)$^b$          & $C_{bb}(\rm H) \times 10^3$ \\
$C_{cc}(\rm H) \times 10^3$ &   $-$1&.10                  &   $-$1&.1~(35)$^b$          & $C_{cc}(\rm H) \times 10^3$ \\
$S_{bb}        \times 10^3$ &   $-$8&.02                  &       &$-$                  & $S_{bb}        \times 10^3$ \\
  \hline
  \end{tabular}\\[2pt]
\tablecomments{$^{a}$ Watson's S reduction in the I$^r$ representation was used in the present fit whereas Watson's A reduction was used in the previous fit \citep{HNSO_IR-1_1996}. Hyperfine parameters were taken from \citet{HNSO_rot_HFS_1993}.  Numbers in parentheses are one standard deviation in units of the least significant figures. Numbers without uncertainties were kept fixed in the fit, see Section~\ref{rot_backgr}. $^{b}$ Derived parameter. \citet{HNSO_rot_HFS_1993} determined $\chi _{aa}$ and $\chi _{bb} - \chi _{bb}$ 
for $^{14}$N and $C_{bb} + C_{cc}/2$ as well as $C_{bb} - C_{cc}$ for both nuclei.}
\end{table*}

The agreement of the rotational and centrifugal distortion parameters from this study and from \citet{HNSO_IR-2_gs-par_1996} is, unsurprisingly, good when the change in reduced Hamiltonian is taken into account. The diagonal quartic distortion parameters $D_J$, $D_{JK}$, and $D_K$ differ from the corresponding $\Delta _J$, $\Delta _{JK}$, and $\Delta _K$ by small multiples of $d_2$; 
$\delta _J = -d_1$, and only the relation between $d_2$ and $\delta _K$ are somewhat more complex. The differences in values are mostly caused by the change in reduction. Similar, albeit more complex relations apply to the sextic distortion parameter to convert from one reduction to the other. Even though our fit contains one distortion parameter less than the fit in \citet{HNSO_IR-2_gs-par_1996}, 
most of their uncertainties are smaller by factors of a few, suggesting their unavailable data set was more extensive or more accurate than the one available to us from \citet{HNSO_IR_2006}. The uncertainties of $B$ and $C$ are essentially the same, which is most likely a consequence of a too low weight of the two hyperfine free transition frequencies of \citet{HNSO_rot_HFS_1993} in the fit of \citet{HNSO_IR-2_gs-par_1996}. The present nuclear quadrupole coupling parameters agree well with those from \citet{HNSO_rot_HFS_1993}; this applies also to the nuclear spin-rotation parameters if we take the relatively large uncertainties from the previous study into account. 

Finally, we provide in Table \ref{t:pfun} the rotational ($Q$$_r$) partition function of the ground state of HNSO. We used SPCAT \citep{Pickett:1991cv} to estimate the values of $Q$$_r$ by direct summation of the ground state energy levels up to $J$ = 125 and $K$$_a$ = 62. These values are provided for the conventional temperatures as implemented in the JPL database \citep{1998JQSRT..60..883P}, and two additional temperatures of 2.725 K and 5.000 K.

\begin{table}
\begin{center}
\caption{Rotational (Q$_r$) partition function of HNSO.}
\label{t:pfun}
\begin{tabular}{cc}
\hline
\multicolumn{1}{c}{Temperature}{\small (K)} & \multicolumn{1}{c}{Q$_r$} \\  
\hline
2.725 &	37.8087 \\
5.000 &	91.8833 \\
9.375 &	233.001 \\
18.75 &	654.456 \\
37.50 &	1844.94 \\
75.00 &	5211.15 \\
150.0 &	14737.8 \\
225.0 &	27087.7 \\
300.0 &	41729.9 \\
\hline 
\end{tabular}
\end{center}
\vspace*{-2.5ex}
\end{table}

\section{Analysis of \ch{NS}}
\label{AnalysisNS}

\restartappendixnumbering

\begin{figure*}
\centerline{\resizebox{0.8\hsize}{!}{\includegraphics[angle=0]{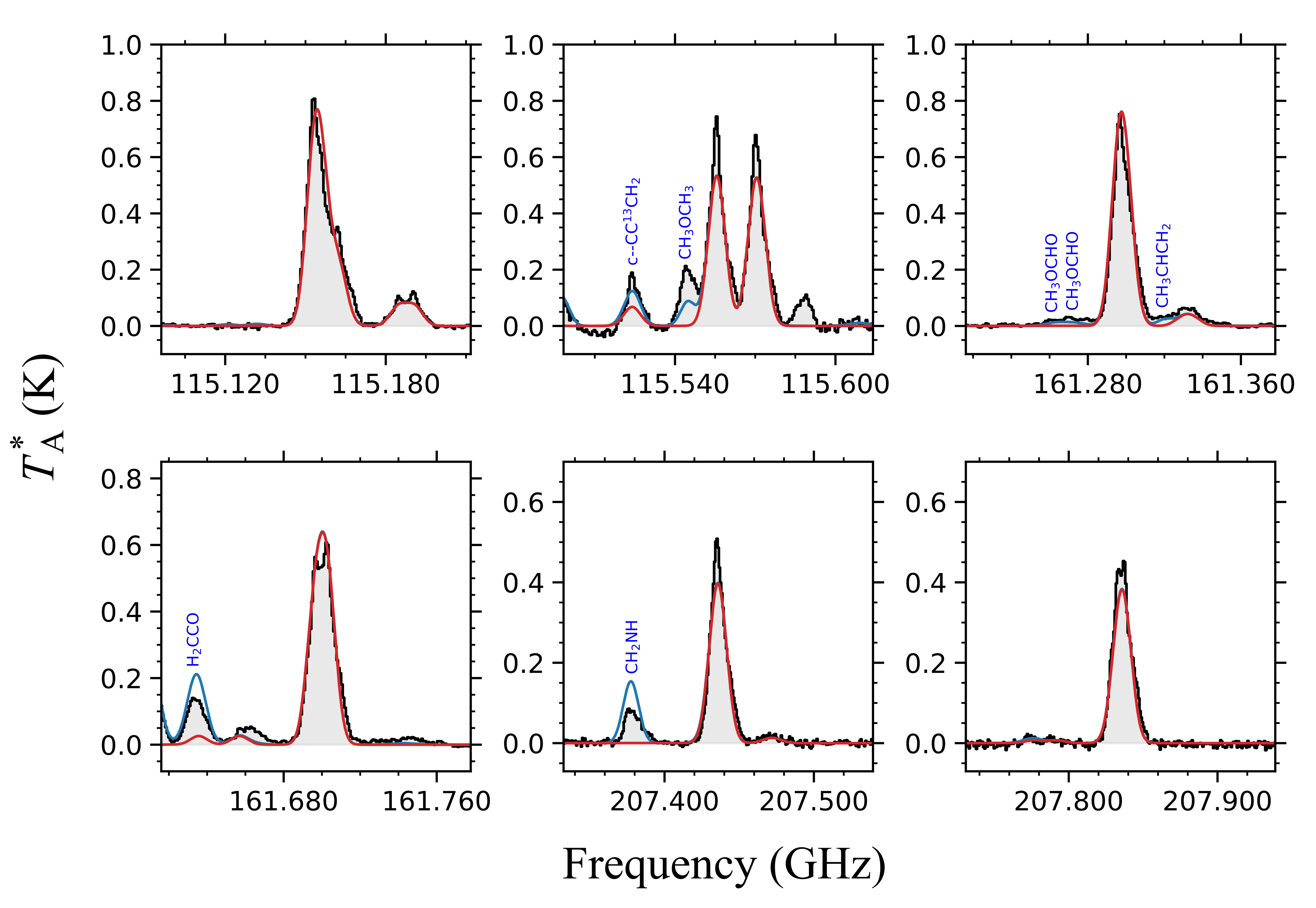}}}
\caption{Unblended or slightly blended transitions of NS detected toward G+0.693 molecular cloud (listed in Table \ref{tab:so2}). The result of the best LTE fit is shown with a red solid line, while the blue line shows the expected molecular emission from all the molecular species identified to date in our survey. The observed spectra are plotted as gray histograms.}
\label{f:ns}
\end{figure*}

For completeness, we also present the LTE analysis of NS ($^2$$\Pi$$_{1/2}$ state), which was performed using all the most unblended transitions (see Table \ref{tab:ns}). We obtained the spectroscopic data from the 046515 entry of the CDMS catalogue \citep{Lee1995}. We present in Figure \ref{f:ns} the result of the best LTE fit using the \textsc{Autofit} tool within SLIM. As shown, the hyperfine structure of NS is partially resolved within the astronomical data set. We derived the following physical parameters from the fit: $N$ = (28 $\pm$ 3) $\times$10$^{13}$ cm$^{-2}$, $T_{\rm ex}$ = (7.5 $\pm$ 0.5) K, $v$$_{\rm LSR}$ = (70.6 $\pm$ 0.6) km s$^{-1}$ and FWHM = (19 $\pm$ 1) km s$^{-1}$. The derived column density is translated into a fractional abundance with respect to H$_2$ of (2.1 $\pm$ 0.2) $\times$10$^{-9}$. In Figure \ref{f:ns}, we depict the fitted line profiles of NS in red, while the expected molecular emission from all the molecules detected to date toward G+0.693 is shown in blue.

\begin{table*}
\centering
\tabcolsep 3pt
\caption{Spectroscopic information of the unblended transitions of NS$^{(a)}$ detected toward G+0.693 (shown in Figure \ref{f:ns}).}
\begin{tabular}{cccccCccccccc}
\hline\hline
Parity & Frequency & $J$' -- $J$'' & $F$' -- $F$'' & log \textit{I} (300 K) & \textit{g}$\mathrm{_u}$ & $E$$\mathrm{_{up}}$ & rms & $\int$ $T$$\mathrm{_A^*}$d$v$ & S/N $^{(b)}$ & Blending  \\ 
& (GHz) &  & &  (nm$^2$ MHz) &   &  (K)  & (mK) & (mK km s$^{-1}$)  & \\
\hline
e & 115.153935 (20) & 2.5 -- 1.5  & 3.5 -- 2.5  & --3.3836  & 8  & 8.8 & 9.1 & 22190 & 425 &  Unblended \\
e & 115.156812 (20) & 2.5 -- 1.5  & 2.5 -- 1.5  & --3.5843  & 6  & 8.8 & & & &  Unblended \\
e & 115.162982 (20) & 2.5 -- 1.5  & 1.5 -- 0.5  & --3.3836  & 4  & 8.8 & & & &  Unblended \\
e & 115.185336 (9) & 2.5 -- 1.5  & 1.5 -- 1.5  & --3.3836  & 4  & 8.8 & 9.1 & 3189 & 61 &  Unblended \\
e & 115.191456 (12) & 2.5 -- 1.5  & 2.5 -- 2.5  & --3.3836  & 6  & 8.8 & & & &  Unblended \\
f & 115.524603 (20) & 2.5 -- 1.5  & 1.5 -- 1.5  & --4.3015  & 4  & 8.8 & 9.1 & 2067 & 50 &  c-CC$^{13}$CH$_2$ \\
f & 115.556253 (20) & 2.5 -- 1.5  & 3.5 -- 2.5  & --3.3807  & 8  & 8.8 & 9.1 & 13159 & 252 &  Unblended \\
f & 115.570763 (20) & 2.5 -- 1.5  & 2.5 -- 1.5  & --3.5814  & 6  & 8.8 & 9.1 & 12290 & 236 &  Unblended \\
f & 115.571954 (20) & 2.5 -- 1.5  & 1.5 -- 0.5  & --3.8067  & 4  & 8.8 &  &  &  &  Unblended \\
e & 161.297246 (20) & 3.5 -- 2.5  & 4.5 -- 3.5  & --2.9737  & 10  & 16.5 & 2.0 & 16153 & 1409 &  Unblended \\
e & 161.298411 (20) & 3.5 -- 2.5  & 3.5 -- 2.5  & --3.1076  & 8  & 16.5 & & & &  Unblended \\
e & 161.301747 (20) & 3.5 -- 2.5  & 2.5 -- 1.5  & --3.2467  & 6  & 16.5 & 2.0 & 2281 & 199 &  \ch{CH3CHCH2}\\
e & 161.330290 (10) & 3.5 -- 2.5  & 2.5 -- 2.5  & --4.1587  & 6  & 16.5 &  &  &  &  Unblended \\
f & 161.636517 (20) & 3.5 -- 2.5  & 3.5 -- 3.5  & --4.1567  & 8  & 16.5 & 2.0 & 3150 & 275 &  \ch{H2CCO} \\
f & 161.657816 (20) & 3.5 -- 2.5  & 2.5 -- 2.5  & --4.1567  & 6  & 16.5 & 2.0 & 1104 & 96 &  Unblended \\
f & 161.697257 (20) & 3.5 -- 2.5  & 4.5 -- 3.5  & --2.9717  & 10  & 16.5 & 2.0 & 16860 & 1471 &  Unblended \\
f & 161.703404 (20) & 3.5 -- 2.5  & 3.5 -- 2.5  & --3.1055  & 8  & 16.5 & & & &  Unblended\\
f & 161.703987 (20) & 3.5 -- 2.5  & 2.5 -- 1.5  & --3.2446  & 6  & 16.5 & & & &  Unblended \\
e & 207.436051 (6) & 5.5 -- 4.5  & 5.5 -- 4.5  & --2.6730  & 12  & 26.4 & 7.5 & 9050 & 211 &  Unblended \\
e & 207.436636 (5) & 5.5 -- 4.5  & 4.5 -- 3.5  & --2.7742  & 10  & 26.4 & & & &  Unblended \\
e & 207.438692 (20) & 5.5 -- 4.5  & 3.5 -- 2.5  & --2.8771  & 8  & 26.4 & & & &  Unblended \\
e & 207.470606 (11) & 5.5 -- 4.5  & 3.5 -- 3.5  & --4.0586  & 8  & 26.4 & 7.5 & 282 & 7 &  Unblended \\
e & 207.475341 (11) & 5.5 -- 4.5  & 4.5 -- 4.5  & --4.0586  & 10  & 26.4 & & & &  Unblended \\
f & 207.834866 (20) & 5.5 -- 4.5  & 5.5 -- 4.5  & --2.6715  & 12  & 26.4 & 7.5 & 9511 & 221 &  Unblended \\
f & 207.838365 (20) & 5.5 -- 4.5  & 4.5 -- 3.5  & --2.7727  & 10  & 26.4 & & & &   Unblended \\
f & 207.838365 (20) & 5.5 -- 4.5  & 3.5 -- 2.5  & --2.8756  & 8  & 26.4 & & & &  Unblended \\					
\hline 
\end{tabular}
\label{tab:ns}
\vspace*{1ex}
\tablecomments{$^{(a)}$ The rotational energy levels are labelled using the quantum number $J$ and $F$. The parity is also indicated. For those hyperfine components that are partially or fully coalesced, we provide the integrated intensity and S/N ratio of the mean observed line rather than the values of each component, which is given only once for each group of transitions. Numbers in parentheses represent the predicted uncertainty associated to the last digits.}
\end{table*}

\section{Analysis of \ch{SO2}}
\label{AnalysisSO2}

\restartappendixnumbering

\begin{figure*}
\centerline{\resizebox{0.8\hsize}{!}{\includegraphics[angle=0]{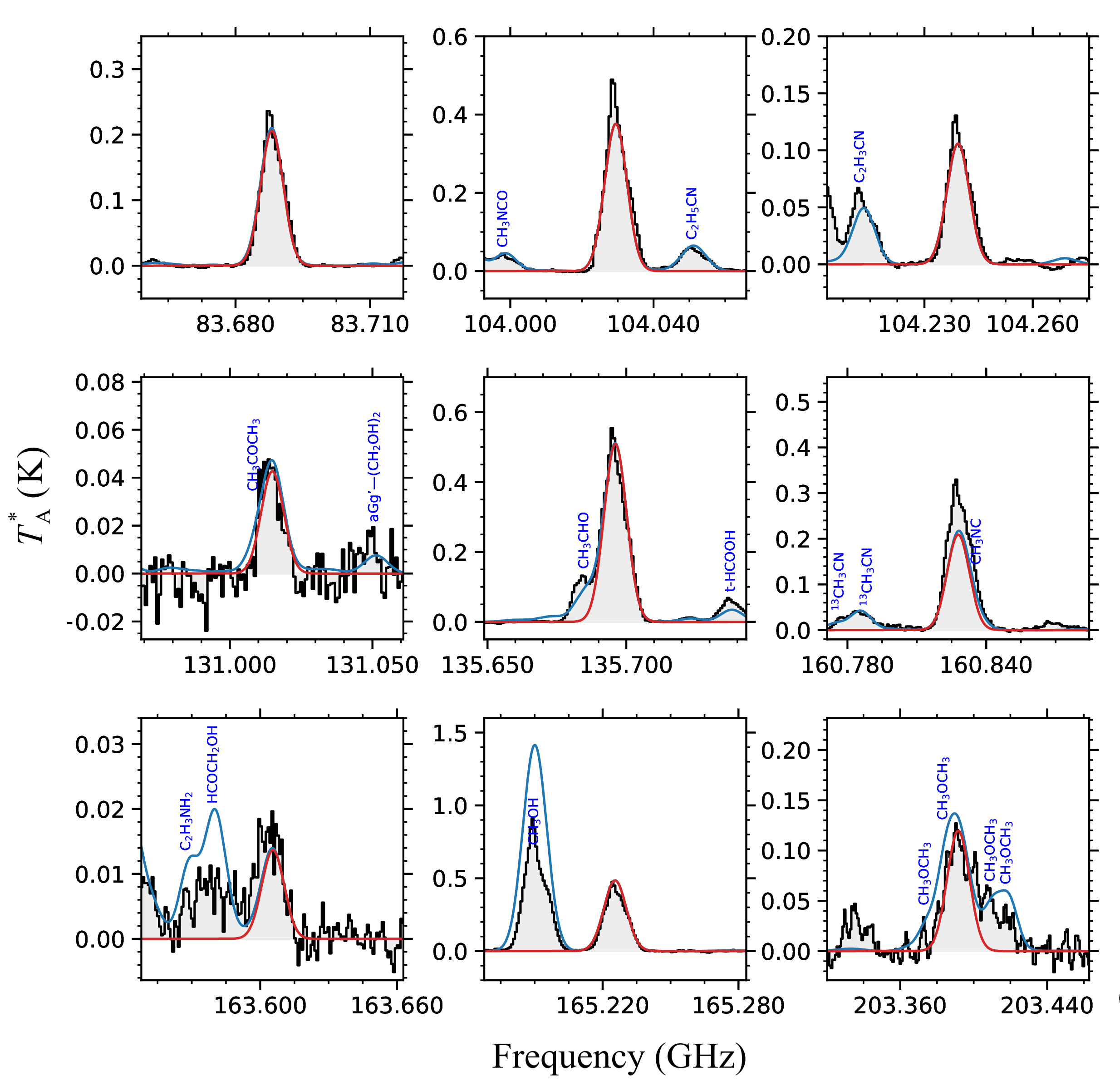}}}
\caption{Unblended or slightly blended transitions of \ch{SO2} detected toward G+0.693 molecular cloud (listed in Table \ref{tab:so2}). The result of the best LTE fit is shown with a red solid line, while the blue line shows the expected molecular emission from all the molecular species identified to date in our survey. The observed spectra are plotted as gray histograms.}
\label{f:so2}
\end{figure*}

We carried out the LTE analysis for \ch{SO2} by using the most intense and unblended transitions clearly detected toward G+0.693 (see Table \ref{tab:so2}). The spectroscopic data were acquired from the 064502 entry of the CDMS catalogue \citep{Muller2005so2}. In Figure \ref{f:so2} we show the result of the best LTE fit derived from \textsc{Autofit}. We obtained the following physical parameters from the LTE fit: $N$ = (38.4 $\pm$ 0.5) $\times$10$^{13}$ cm$^{-2}$, $T_{\rm ex}$ = (18.9 $\pm$ 0.2) K, $v$$_{\rm LSR}$ = (68.9 $\pm$ 0.1) km s$^{-1}$ and FWHM = (21.1 $\pm$ 0.3) km s$^{-1}$. Thus, we derived a fractional abundance with respect to molecular hydrogen of (2.8 $\pm$ 0.3) $\times$10$^{-9}$. We used the same color code for the line profiles of Figure \ref{f:so2} as that of Fig. \ref{f:LTEspectrum} (i.e., fitted line profiles of \ch{SO2} in red and the expected molecular emission from all the molecules detected to date toward G+0.693 in blue).

\begin{table*}
\centering
\caption{Spectroscopic information of the selected transitions of \ch{SO2} detected toward G+0.693 (shown in Figure \ref{f:so2}).}
\begin{tabular}{ccccccccccc}
\hline
Frequency & Transition $^{(a)}$ & log \textit{I} (300 K) & \textit{g}$\mathrm{_u}$ & $E$$\mathrm{_{up}}$ &  rms & $\int$ $T$$\mathrm{_A^*}$d$v$ & S/N $^{(b)}$ & Blending  \\ 
(GHz) & &  (nm$^2$ MHz) &   &  (K)  & (mK) & (mK km s$^{-1}$)  & \\
\hline
83.6880930 (20)  &  8$_{1,7}$--8$_{0,8}$ & -3.9576  & 17  & 36.5 & 1.3 & 4778 & 641 &  Unblended \\
104.0294183 (20) &  3$_{1,3}$--2$_{0,2}$ & -4.2264  & 7  & 2.7 & 1.2 & 9547 & 1388 &  Unblended \\
104.2392952 (20) &  10$_{1,9}$--10$_{0,10}$ & -3.7708  & 21 & 54.3 & 1.2  & 2619 & 381 &  Unblended \\
131.014860 (80) &  12$_{1,11}$--12$_{0,12}$  & -3.6063  & 25  & 75.9 & 5.4 & 831 & 27 &  \ch{CH3NC} \\
135.696020 (80) &  5$_{1,5}$--4$_{0,4}$  & -3.8143  & 11  & 15.6 & 1.6 & 12587 & 1373 &  Unblended \\
160.827880 (80) &  10$_{0,10}$--9$_{1,9}$  & -3.4028  & 21  & 49.4 & 2.1 & 7239 & 602 &  \ch{CH3OCH3}\\
163.6055328 (6) &  14$_{1,13}$--14$_{0,14}$  & -3.4644  & 29  & 101.0 & 1.6 & 430 & 47 &  Unblended \\
165.2254511 (8) &  7$_{1,7}$--6$_{0,6}$  & -3.5083  & 15  & 26.9 & 1.6 & 9703 & 1058 &  Unblended \\
203.39155 (10) &  12$_{0,12}$--11$_{1,11}$  & -3.1094  & 25  & 69.6 & 9.3 & 3668 & 69 &  Unblended \\
\hline 
\end{tabular}
\label{tab:so2}
\vspace*{1ex}
\tablecomments{$^{(a)}$ The rotational energy levels are labelled using the conventional notation for asymmetric tops: $J_{K_{a},K_{c}}$, where $J$ denotes the angular momentum quantum number, and the $K_{a}$ and $K_{c}$ labels are projections of $J$ along the $a$ and $c$ principal axes. Numbers in parentheses represent the predicted uncertainty associated to the last digits.}
\end{table*}

\end{document}